\documentclass[twocolumn]{aastex631}

\begin{document}

\title{An image-based blind search for Fast Radio Bursts in 88 hours of data from the EoR0 Field, with the Murchison Widefield Array}

\correspondingauthor{Ian Kemp}
\email{ian.kemp@icrar.org, ian.kemp@postgrad.curtin.edu.au}

\author[0000-0002-6637-9987]{Ian Kemp}
\affiliation{International Centre for Radio Astronomy Research (ICRAR) \\
Curtin University, Bentley, WA 6102, Australia}
\affiliation{CSIRO Space and Astronomy, \\ 
26 Dick Perry Avenue, Kensington, WA 6151, Australia}

\author[0000-0002-8195-7562]{Steven Tingay}
\affiliation{International Centre for Radio Astronomy Research (ICRAR) \\ 
Curtin University, Bentley, WA 6102, Australia}

\author{Stuart Midgley}
\affiliation{Defence Science and Technology Group \\
Australian Department of Defence, Australia.}

\author[0000-0002-1828-1969]{Daniel Mitchell}
\affiliation{CSIRO Space and Astronomy \\
Cnr Vimiera \& Pembroke Roads
Marsfield NSW 2122, Australia}

\begin{abstract}

This work is part of ongoing efforts to detect Fast Radio Bursts (FRBs) using the Murchison Widefield Array (MWA) in a spectral window below 300 MHz. We used an image-based method based on the pilot study of Tingay et al. 2015, scaled up via massively parallel processing using a commercial supercomputer.  We searched 87.6 hours of 2-second snapshot images, each covering 1165 square degrees of the EoR0 field, over a dispersion measure range of 170 to 1035 pc cm\textsuperscript{-3}.  The large amount of data necessitated the construction of a series of filters to classify and reject the large number of false positives. Our search was more sensitive than any previous blind search using the MWA telescope, but we report no FRB detections, a result which is consistent with the extrapolation into the low-frequency domain of the results of \citet{2024arXiv240104346S}. We obtain upper bounds on the event rate ranging from $<$1783 sky\textsuperscript{-1}day\textsuperscript{-1} at a fluence of 392 Jy ms, to $<$31 sky\textsuperscript{-1}day\textsuperscript{-1} at 8400 Jy ms, for our spectral window of 167-198 MHz.  Our method was shown to be computationally efficient and scalable by the two or three orders of magnitude required to seriously test the model of Sokolowski et al. Our process is especially sensitive to detections of satellites and meteor trails and may find applications in the identification of these transients. We comment on future surveys using this method, with both the MWA and the SKA.
\end{abstract}

\keywords{Radio transient sources (2008) --- Radio astronomy (1338) --- Computational methods(1965)}

\section{Introduction} \label{sec:introduction}

Fast Radio Bursts (FRBs) are highly energetic bursts of energy at radio wavelengths, typically with durations up to a few milliseconds. They are believed to originate from cosmological sources \citep{2019A&ARv..27....4P}, but the actual mechanism(s) producing them are still unknown. Therefore FRBs are an active subject of research by several radio astronomy groups worldwide.  Since the first detection of an FRB in 2007 \citep{2007Sci...318..777L}, searches to 17 May 2024 yielded 941 detections listed on the Transient Name Server \citep{2020TNSAN.160....1P}, with over 2/3 of them discovered by the CHIME telescope in Canada \citep{2018ApJ...863...48C} and ASKAP in Australia \citep{2010PASA...27..272M}. It is known that some sources of FRBs are `repeaters', originating multiple bursts; but the majority apparently generate one-off events \citep{2019A&ARv..27....4P}. 

There is circumstantial evidence linking FRBs with observed radio bursts from magnetars, although several other hypotheses for their origin exist \citep{2022A&ARv..30....2P}. In order to help elucidate their origin mechanism, there is significant value in detecting and characterising FRBs, localising them if possible, and in particular collecting data in frequency ranges where observations are scarce \citep{2022A&ARv..30....2P}.  The objective of the present work is to execute a large scale `blind' search, using the MWA telescope (\citet{2009IEEEP..97.1497L}, \citet{2013PASA...30....7T}, \citet{2018PASA...35...33W}) in the range 167 to 198 MHz. We use an image-based method piloted by \citet{2015AJ....150..199T} and leverage a large commercial computational resource to search historical observations with a larger parameter space than used in the pilot study.

Of the 941 FRBs discovered to date, all but one FRB have been detected in the mid-to high frequency range, above 350 MHz.  Since 2018, the large number of detections using the CHIME telescope in the range 400–800 MHz \citep{2018ApJ...863...48C} have confirmed an apparent random sky distribution of FRBs, an estimate of the rate of $<$820 sky\textsuperscript{-1}day\textsuperscript{-1} at a fluence of $>$5 Jy ms at 600 MHz, and support for a -3/2 power law relation between fluence and dispersion measure (DM) \citep{2021ApJS..257...59C}, \citep{2023ApJS..264...53C}.

Previous low frequency FRB searches ($\lesssim$850 MHz) were recently reviewed by \citet{2021Univ....8....9P}.  Below 300 MHz, searches using Arecibo \citep{2013ApJ...775...51D}, the Lovell Telescope \citep{2020MNRAS.493.4418R}, and LOFAR (\cite{2019A&A...623A..42H}, \cite{2014A&A...570A..60C}, \cite{2015MNRAS.452.1254K}, \cite{2020MNRAS.491..725M}) provided no detections, and we currently have one single detection of a repeater by LOFAR, reported by \citet{2021ApJ...911L...3P} and \citet{2021Natur.596..505P}.

Prospects for FRB detection with the MWA were estimated by \citet{2013ApJ...776L..16T}. These authors used the known characteristics of the telescope, and extrapolated the then-known source population based on the (unknown) spectral index, $\alpha$, considering both high and low scattering scenarios. For observations in imaging mode, the authors estimated a rate of between 0.2 ($\alpha$=0.2) to 38 ($\alpha$=-2) 7-$\sigma$ detections per 10 hours of observation.

A more recent estimate by \citet{2024arXiv240104346S} extrapolates from detections in existing mid-high frequency searches. This method allows an estimate of detection rates for the low-sensitivity image based method, and will be used to interpret results from the present work. The model suggests an event rate of greater than 100 sky\textsuperscript{-1}day\textsuperscript{-1} at a fluence greater than 100 Jy ms should be detectable with the MWA.

\begin{deluxetable*}{lccccccc}[th]
\tablecolumns{8}
\tablecaption{FRB searches using MWA. For each of the four searches, this table lists the total
observation time searched, the observation frequency, the time and frequency resolution, the range of dispersion measures searched, the sensitivity, and the bounds obtained on event rate and spectral index.} \label{table:MWAsearches}
\tablehead{
\colhead{Reference} & \colhead{Obs. Time} & \colhead{Freq.} & \colhead{Time/Freq} & \colhead{DM} & \colhead{Sensitivity} & \colhead{Event rate} & \colhead{$\alpha$} \\
 & \colhead{(hr)} & \colhead{(MHz)} & \colhead{resolution} & \colhead{(pc cm\textsuperscript{-3})} & \colhead{(Jy ms)} & \colhead{(sky\textsuperscript{-1} day\textsuperscript{-1})} & 
}
\startdata
Tingay et al. 2015 & 10.5 & 156 & 2 s/1.28 MHz & 170-675 & 700 & $<$ 700 & $>$ -1.2 \\ 
Rowlinson et al. 2016 & 100 & 182 & 28 s/30.72 MHz & $<700\dagger$ & 7980 & $<$ 82 & $>$ -1.0 \\ 
Sokolowski et al. 2018 & 3.5 & 185 & 0.5 s/1.28 MHz & 343-715 & 450-6500 & & \\
Tian et al. 2023 & 24.1 & 144-215 & 400 $\mu$s/10 kHz & 411-651 & 32-1175 & & \\
& & & & \multicolumn{4}{l}{$\dagger$ Estimated by the current authors}\\
\enddata
\end{deluxetable*} 

To date there have been four FRB searches using the MWA. These have employed both high time resolution, using the Voltage Capture System (VCS) \citep{2015PASA...32....5T}, and blind image searches - summarised in Table \ref{table:MWAsearches} (adapted from \cite{2023MNRAS.518.4278T}). These four searches returned no FRB detections.

The search by \citet{2015AJ....150..199T} was based on searching the dynamic spectra of pixels in cleaned and model-subtracted 2-second snapshots. This process obtained an RMS noise of 50 mJy/beam. Over one hundred 6$\sigma$ detections were investigated manually and rejected as FRBs, leading to an upper bound on the event rate, of 700 sky\textsuperscript{-1} day\textsuperscript{-1} at 700 Jy ms. The computing cost for the search was approximately 36 hours per hour of observation using a single CPU.

\citet{2016MNRAS.458.3506R} analysed 100 hr of observations of the same EoR0 field, but at a lower sensitivity. Images using 28 second snapshots were cleaned and source-subtracted, giving an RMS noise of 31.4 mJy/beam.  No de-dispersion was carried out. Three $6\sigma$ detections were made and identified as artefacts.  The upper bound event rate was estimated as 82 sky\textsuperscript{-1} day\textsuperscript{-1} at 700 Jy ms. The computational load of the search was not reported, but the authors imply that computational scale limited their search to 100 hrs of EoR observations.

\citet{2018ApJ...867L..12S} used the MWA to shadow the CRAFT transient search at higher frequency using the Australian Square Kilometre Pathfinder (ASKAP) (See \citet{2021PASA...38....9H} for details of this telescope, \citet{2010PASA...27..272M} for details of CRAFT), but no simultaneous detections were made of any of the 7 FRBs observed by ASKAP.

\citet{2023MNRAS.518.4278T} used high time-resolution data from the MWA VCS \citep{2015PASA...32....5T}, in a targeted search. A search of 23.3 hours of MWA archived data at the known locations of repeaters detected by other telescopes also yielded no FRB candidates.

Our study follows on from these searches, by scaling up the searched parameter space. As we move to larger searches, we expect correspondingly more false positives, for which the manual inspection methods used in earlier searches will be impractical. Therefore we wish to improve the detection and classification of these false positives.

Previous work has shown that false positives can result from several distinct sources. \citet{2015AJ....150..199T} and later work have shown that high cadence imaging is sensitive to the effects of interplanetary scintillation (IPS), to the extent that it can be used for detailed studies of space weather (for example \citet{2023SpWea..2103396M}). 

\citet{2015PASA...32....8O} noted several types of radio frequency interference (RFI) at the MWA, including short bursts of broadband interference of unidentified origin, narrow band interference corresponding to the 2 m amateur radio band, and also broadcast digital TV (DTV).  Using the Engineering Development Array (EDA2) adjacent to the MWA, \citet{2020PASA...37...39T} were able to classify the origins of narrowband transients (at FM frequencies) as reflections from aircraft, satellites, and meteor trails, as well as signals direct from the transmitters, propagated over the horizon by tropospheric scattering.

A further source of RFI is emission from satellites. \citet{2020PASA...37...10P} and \citet{2023A&A...678L...6G} have detected radio emission from a number of satellites using the MWA and EDA2, respectively. Such transmissions are detected as either broadband, or multiple-narrowband detections, and arise from both intentional transmission and unintentional emission from onboard electronics.

The key component of our large scale FRB search has been to develop improved automated and statistical methods to deal with these sources of false positives. In the following section we explain the processing methods developed to achieve scale up using massively parallel processing.

\section{Observations and Data Processing}
\subsection{Observations}
The data set considered is a collection of 3348 files, each containing approximately 2 minutes of visibility data, totalling 107.1 hours of observations of the EoR0 field centered 
at $\alpha$ = 00$^h$ $\delta$ = -27\textdegree \hspace{1pt} \citep{2016ApJ...819....8P}. 2027 observations were of 112 second duration and 1321 were of 120 second duration. 

Observations were obtained on 127 distinct days between 14/08/2013 and 08/12/2021, and are listed in Table \ref{table:MWA_observations}. This was a commensal search, in that the data were originally collected for EoR research. 3014 of the observations had been extracted from the MWA data archive by the EoR team, in 2022, for error-checking and calibration using the
`Hyperdrive'\footnote{\url{https://github.com/MWATelescope/mwa_hyperdrive}} software package developed by the EoR team.

In addition to these data, the 334 observations searched by Tingay et al in 2015 were reprocessed, and this work was used to develop our processing pipeline. These observations were calibrated using cotter \citep{2015PASA...32....8O}.

All observations were made with a single spectral window of 167 to 198 MHz. For compatibility with earlier searches, data were frequency-averaged into 24x1.28 MHz channels.  

The 3348 observations included 1504 (45\% of the total) collected with the array in the Phase I configuration \citep{2013PASA...30....7T} and 1844 (55\%) in the Phase II compact configuration \citep{2018PASA...35...33W}. The maximum baselines for the two configurations were 2438 m and 750 m, respectively. Phase I data were processed to images of 1024x1024 pixels with 2 arcmin pixel size, Phase II data were processed to 256x256 pixels with 8 arcmin pixel size.

Accounting for differences in calibration and array configuration the observations fall into three categories (see Table \ref{table:observations}).

\begin{deluxetable*}{clllllll}
\tablecolumns{8}
\tablecaption{Summary of Observations and Images}
\label{table:observations}
\tablehead{
\colhead{Category} & \colhead{Description} & \colhead{MWA Telescope} & \colhead{Calibration} & \colhead{Count} & \colhead{Duration} & \colhead{Image Size} & \colhead{Image} \\
 &  & \colhead{Configuration} & & & \colhead{(hr)} & (pixels) & \colhead{Resolution} 
}
\startdata
A & Tingay et al. 2015 & Phase I Extended & Cotter & 334 & 10.4 & 1024x1024 & 2 arcsec \\
B & EoR extended & Phase I Extended & Hyperdrive & 1170 & 36.4 & 1024x1024 & 2 arcsec \\
C & EoR compact & Phase II Compact & Hyperdrive & 1844 & 60.3 & 256x256 & 8 arcsec \\ 
 \hline
 & TOTAL & & & 3348 & 107.1 & & \\ 
\enddata
\end{deluxetable*}
\begin{deluxetable*}{cccc|cccc}
\tablecolumns{8}
\tablecaption{Sample of MWA Observations. This table lists the MWA observations used in the current study - start date, start and stop times, all UTC}
\label{table:MWA_observations}
\tablecomments{Table 3 is published in its entirety in the machine-readable format. A portion is shown here for guidance regarding its form and content.}
\tablehead{
\colhead{Obs. Id} & \colhead{Start Date} & \colhead{Start Time} & \colhead{Stop Time} & \colhead{Obs. Id} & \colhead{Start Date} & \colhead{Start Time} & \colhead{Stop Time}
}
\startdata
\multicolumn{4}{c|}{Category A: Phase 1 Extended array config} & \multicolumn{4}{c}{Category C - Phase II Compact array config (cont.)} \\
1096809200 & 8/10/14 & 13:13:04 & 13:14:56 & 1165836936 & 15/12/16 & 11:35:19 & 11:37:11 \\
1096809320 & 8/10/14 & 13:15:04 & 13:16:56 & 1185484136 & 30/07/17 & 21:08:38 & 21:10:30 \\
1096809440 & 8/10/14 & 13:17:04 & 13:18:56 & 1185484248 & 30/07/17 & 21:10:30 & 21:12:22 \\
1096809560 & 8/10/14 & 13:19:04 & 13:20:56 & 1185484360 & 30/07/17 & 21:12:22 & 21:14:14 \\
\enddata
\end{deluxetable*}

\subsection{Computational strategy}
In our work, the basic unit of computation was a two-minute observation, corresponding to individual observation IDs (Table \ref{table:MWA_observations}). Observations were processed into sequences of two-second snapshot image cubes, which were handled together based on their parent observation. Statistics were calculated on observations, as well as on individual snapshot images. Processing was carried out using commercial facilities provided by DUG Technology in Perth, Western Australia\footnote{\url{https://dug.com/about-dug/}}. The key consideration was to provide an efficient, low cost search, at large scale. It was found that this was best achieved by allocating one 2-minute observation per node for processing and searching, minimising traffic between the node and the file system. We chose to use nodes consisting each of a single low-cost Intel Xeon Phi `KNL' processor, with each node provisioning 64 cores and 128 GB of memory. Allocation of one 2-second snapshot per core permitted all snapshots in an observation to be searched concurrently. Part of the node memory was configured as `RAM drive' to store the visibility data, temporary files, and images, with the remainder used to carry out de-dispersion and searching, via array processes wholly in memory. This resulted in an average run time of 15 minutes for imaging and search of a 2-minute observation, which is satisfactory compared with a serial disk-based computation which was found to consume 72 node-hours.  The approach was more amenable to parallel processing than the method used in the 2015 pilot, which required observations to be assembled in sequence. The trade off is that the method produces a bias towards lower DM values, which will be discussed fully in section \ref{sec:dedispersion}.  With thousands of KNL nodes available at the DUG facility, the processing of a batch of 400 2-minute observations was typically completed in under 25 minutes.

\subsection{Imaging} \label{sec:imaging}
Visibilities were inverted into `dirty' 2-second snapshot image cubes using WSClean \citep{offringa-wsclean-2014}. Snapshots using XX and YY polarisations were produced and processed separately.

Successive 2s x 24 x 1.28 MHz snapshots were differenced to remove persistent sources and sidelobes in the non-de-dispersed domain, while retaining short period transients. A Multi-Frequency Synthesis (MFS) image was computed for each differenced snapshot and Gaussian statistics calculated. A typical dirty image cube calibrated via hyperdrive, for a single polarisation, had an RMS of 315 mJy/beam in the central channel, and the corresponding 2 s x 30.72 MHz MFS image had an RMS of 76 mJy/beam. After differencing, the RMS values would typically be 204 mJy/beam for the central channel and 42 mJy/beam for the MFS image.  Differenced images with RMS less than 50 mJy/beam or more than 1000 mJy/beam in the central channel of the corresponding cube were flagged as `blank' or `RFI contaminated' and excluded from further processing.

This process identified extensive RFI in six 2-minute observations, which were then excluded from processing. In the remaining data, the flagging process mainly identified empty images near the start or end of the observation, with only 21 mid-observation images excluded.

\begin{figure}[h]
\plotone{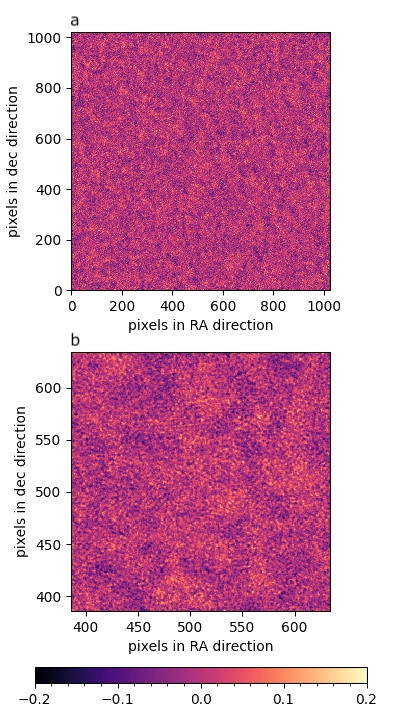}
\caption{Typical (non-de-dispersed) MFS differenced image of a snapshot from Category B, XX polarisation. a = full image; b = central portion, showing texture. Scale units are Jy/beam.} 
\label{fig:Example_normal_snapshot_full}
\end{figure}

A typical MFS differenced image from Category B (as defined in Table \ref{table:observations}) is shown in Figure \ref{fig:Example_normal_snapshot_full}a. The central portion of the same image is shown in Figure \ref{fig:Example_normal_snapshot_full}b, and illustrates a random texture with characteristic scale of around 50 pixels (100 arcsec), which was found to vary with time during the observation and vary in amplitude between observations.  This texture was attributed to aliasing of the setting galactic plane, recently modelled by \citet{2024ApJ...964..158B}.

\subsection{IPS and Satellite rejection} \label{sec:IPS}
\begin{figure}
\plotone{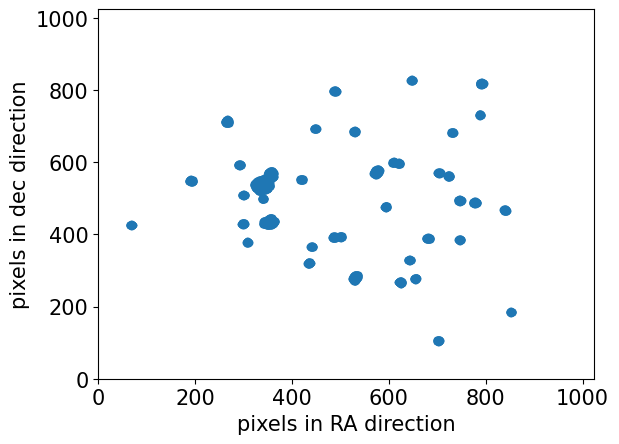}
\caption{Example Scintillation Mask 
\label{fig:Scint_Mask}}
\end{figure}

Interplanetary scintillation (IPS) was dealt with by processing images in batches of 400 2-minute observations ($\sim$48,000 2-second snapshots). A `scintillation mask' was built containing all pixels in the batch which reported $> 6\sigma$ in six or more separate differenced MFS images, and their neighbours within 10 arcmin. The threshold of six images was adopted to avoid inadvertently removing true transients. Neighbours were included to remove most anomalous pixels $<6\sigma$. This process caused 1.5$\%$ of the data to be excluded (mostly at bright persistent sources and their sidelobes). One of the scintillation masks is shown in Figure \ref{fig:Scint_Mask}. For efficiency, the mask was built using pixel coordinates rather than sky coordinates. This choice meant that we could not use the small amount of driftscan data available in the overall set of observations (Table \ref{table:pre-processing}). 

\begin{deluxetable}{rl}
\tablecolumns{2}
\tablecaption{Observations excluded from search}
\label{table:pre-processing}
\tablehead{
\colhead{Count} & \colhead{Description}
}
\startdata
3348 & Total calibrated observations \\ 
 \hline
207 & Drift-scan data \\
6 & Extensive RFI \\ 
8 & Contain satellite tracks \\ 
 \hline
3127 & Suitable for searching \\
\enddata
\end{deluxetable}

\begin{deluxetable*}{lcccc}
\tablecolumns{5}
\tablecaption{Satellites detected in the parent data set}
\label{table:satellites}
\tablehead{
\colhead{Obs. id} & \colhead{Timestep} & \colhead{Time Range (UTC)} & \colhead{Name} & \colhead{NORAD \#}
}
\startdata
1062522960 & 49-51 & 06-Sep-2013/17:15:41.0 - 17:17:33.0 & \multicolumn{2}{c}{Unidentified} \\	
1066572616 & 15-27 & 23-Oct-2013/14:09:58.0 - 14:11:48.0 & ALOS & 28931 \\
1087596040 & 06-39 & 23-Jun-2014/22:00:24.7 - 22:02:10.7 & \multicolumn{2}{c}{Unidentified} \\
1088806248 & 04-40 & 07-Jul-2014/22:10:32.7 - 22:12:18.7 & \multicolumn{2}{c}{Unidentified} \\
1094751016 & 12-19 & 14-Sep-2014/17:30:00.7 - 17:31:46.7 & \multicolumn{2}{c}{Unidentified} \\
1160576968 & 25-37 & 15-Oct-2016/14:29:10.7 - 14:30:58.7 & \multicolumn{2}{c}{Unidentified} \\
1096811152 & 35-45 & 08-Oct-2014/13:45:37.0 - 13:47:29.0 & Ukube 1 & 40074\\
1096812488 & 06-23 & 08-Oct-2014/14:07:53.0 - 14:09:45.0 & Triton-1 & 39427U\\
\enddata
\end{deluxetable*}

\begin{figure}[h]
\plotone{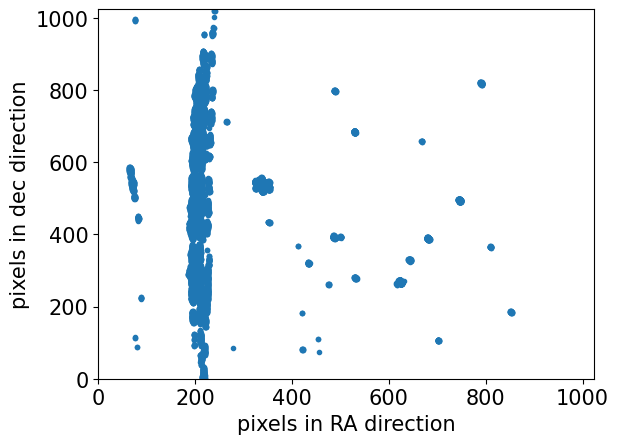}
\caption{Example satellite track (obs. id 106672616), showing all pixels $>6\sigma$ in one 2-minute differenced MFS image set 
\label{fig:Sat_Track}}
\end{figure}

\begin{figure}[h]
\begin{interactive}{animation}{images/1160576968.m4v}
\plotone{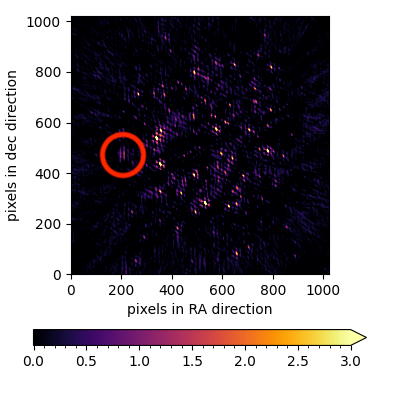}
\end{interactive}
\caption{Satellite image (obs. id 1160576968) \textit{This figure is available as an animation - this 4-second animation shows three cycles of a satellite detection passing through the dirty image from top to bottom near RA = 1hr - the circled feature in the static image is the satellite detection near coordinates (200, 500).} Colour bar scale is Jy/beam}
\label{fig:satellite_image}
\end{figure}

A count was made of the total number of pixels greater than 6$\sigma$ in 2-second snapshot images from each 2-minute observation. Observations with more than 10x the count expected from Gaussian statistics were flagged, and this count was found to be a good diagnostic for the presence of emission from satellites. Plots of the $>6\sigma$ pixels readily showed up arcs or linear tracks - an example is shown in Figure \ref{fig:Sat_Track}. These were confirmed to originate from satellites by constructing movies using the affected MFS images (example: Figure \ref{fig:satellite_image}). In order to identify candidate satellites, we used the times and celestial coordinates of the detected events (Table \ref{table:satellites}) to search for satellites coincident on the sky (within one degree), as seen from the location of the MWA.  We obtained Two Line Element (TLE) orbital element information from space-track.org, for all cataloged satellites at the time of observation, selecting TLEs within two days of the observation date.  For observations a decade ago, the TLE database was not as complete as it is now, and it is possible that TLE updates were not performed at a cadence that would fit within a four day window around the observation for all objects.  However, TLEs more than a few days old rapidly go out of date.  Hence, while we identify strong candidates for a few of our detections, the majority remain unidentified.

The processing yielded 3127 observations (totalling 100.0 hours) suitable for the transient search - the reasons for rejection are summarised in Table \ref{table:pre-processing}. 

\subsection{De-dispersion} \label{sec:dedispersion}
\begin{figure}[h]
\plotone{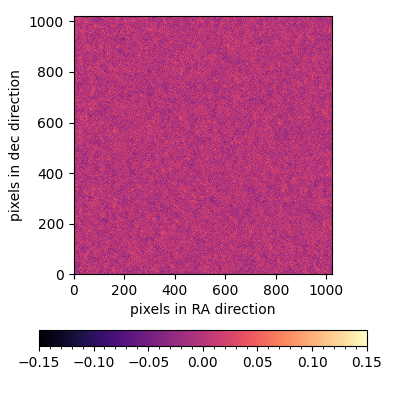}
\caption{Typical (de-dispersed) MFS differenced image from Category B - average of XX and YY polarisations. Colour bar scale is in Jy/beam} 
\label{fig:Example_normal_snapshot_final}
\end{figure}

De-dispersion was achieved by averaging together suitably time-shifted frequency slices from differenced image cubes. To ensure processing fully within the available memory on the supercomputer nodes, we limited the number of trial DM values to 24, ranging from 170 to 1035 pc cm\textsuperscript{-3}, a range which contains 80\% of confirmed FRBs \citep{2019A&ARv..27....4P}. Values were chosen to avoid overlap given the 2-second time resolution of the snapshot images. XX and YY image cubes were processed independently, then averaged together to provide pseudo-Stokes I (non primary-beam corrected) de-dispersed images. This procedure was carried out on the basis that, although most FRB sources are known to be linearly polarised, Faraday rotation leads to the direction of polarisation being effectively random and independent in each frequency band, effectively providing a `rotation measure' (RM) of zero in our spectral window \citep{2022Sci...375.1266F}. An example final de-dispersed, polarisation-averaged MFS image is shown in Figure \ref{fig:Example_normal_snapshot_final}.

The final RMS noise in the snapshot images depended upon the category of the observations, and values are listed in Table \ref{table:search}.

\begin{deluxetable}{cclc}[h]
\tablecolumns{4}
\tablecaption{Summary of FRB search (categories as in Table \ref{table:observations})}
\label{table:search}
\tablehead{
\colhead{Category} & \colhead{No. Snapshots} & \colhead{Duration} & \colhead{Median RMS} \\
 & & \colhead{(hr)} & \colhead{noise (mJy/beam)}
}
\startdata
A & 14,923 & 8.3 & 48 \\
B & 58,581 & 32.6 & 28 \\
C & 84,113 & 46.7 & 60 \\
\hline
TOTAL &  & 87.6 &  \\
\enddata
\end{deluxetable}

Parallel processing the data as individual 2-minute observations leads to a bias towards lower DM values. With our spectral window, the lowest trial DM value uses data from four timesteps, so for a 120 s observation, images can be created from 56 timesteps. The highest DM value uses 22 timesteps, so only 38 can be created. For this reason, the event rate sensitivity in this study varies with DM.

\subsection{Event detection and classification} \label{sec:detection}
A key component of our blind search is the efficient identification and classification of false positives. In our process we extract 7-$\sigma$ signals, and apply a sequence of tests to classify them.

Each de-dispersed image was masked using the scintillation mask, and the mean and RMS intensity was calculated on the remaining pixels, along with an 80-bin histogram which was used to confirm conformance to a Gaussian distribution.  Assuming the noise is uniformly distributed, the RMS value was used as an estimate for $\sigma$, the standard deviation of the underlying distribution. For each image the pixel with the highest intensity was recorded, and data for all images containing pixels $>7\sigma$ were extracted.

Pre-processing and de-dispersion provided a total of 3.09 million de-dispersed images.  The search identified 1519 images with $7\sigma$ detections (0.05\% of the initial image population).

\begin{figure}[h]
\plotone{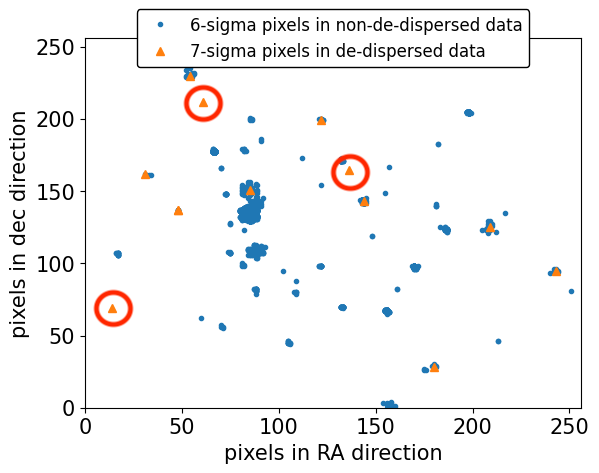}
\caption{Example of screening for repeat non-dispersed signals. Single 7$\sigma$ detections (circled) were retained, others discarded
\label{fig:screening}}
\end{figure}

The sequential tests are detailed below, and the numbers classified at each stage are summarised in Table \ref{table:detections}.
\begin{itemize}
\item Where there were multiple detections at the same time and location, but for different DM values, the highest-$\sigma$ detection was selected and the others discarded. Multiple detections at the same location but at different times were retained, unless they were in adjacent 2-second snapshots. This removed 1,323 detections;
\item 101 detections at a location with two or more $>$6$\sigma$ detections in the corresponding non-de-dispersed data were discarded as artefacts, probably due to residual IPS, see Figure \ref{fig:screening};
\item 21 de-dispersed images were discarded after manual inspection, as the detection lay in an obvious structure, such as a sidelobe which had survived image differencing;
\item Waterfall plots (intensity as a function of de-dispersed time and frequency) were constructed for the detection pixels, and for pixels five steps in each direction from the detection (in pixel coordinates). 61 detections were discarded because the signal was discernable in one or more of the 5-pixel neighbours, as a true FRB detection would appear in a single pixel only (unless strong enough to present sidelobes);
\item A further 6 detections were discarded where the waterfall plot revealed broadband RFI;
\item De-dispersed detections were imaged over a range of pixel sizes (108 to 600 arcsec) and image sizes (256x256 pixels to 2048x2048 pixels). Detections were discarded unless they recorded $>$7$\sigma$ in a majority of the images. This step was carried out to remove detections where the intensity is increased by aliasing of sources outside the image area, and accounted for 4 detections;
\item De-dispersed XX, YY, and I waterfall plots of the remaining 3 detections were examined to check for FRB-like characteristics, and revealed narrowband RFI.
\end{itemize}

The tests accounted for all the 7-$\sigma$ detections, meaning that no FRB candidates were identified in the data. This will be used in the next section to calculate upper bounds on the whole sky event rate in our frequency range.  An important result from the search is that we have developed an efficient pipeline which can be scaled up to larger data sets without incurring excessive manual costs in dealing with false positives.

\begin{deluxetable}{rl}[h]
\tablecolumns{2}
\tablecaption{Classification of 7-$\sigma$ detections}
\label{table:detections}
\tablehead{
\colhead{Count} & \colhead{Classification}
}
\startdata
3,092,490 & Total de-dispersed images \\
1,519 & Total images with 7-$\sigma$ detections \\
\hline
1,323 & Event at multiple DM values \\
101 & Coincides with repeating non-dispersed signal \\
21 & In large scale structure in the de-dispersed image \\
61 & Detection in waterfall plot spans $>$5 pixels \\ 
6 & Broadband RFI identified in waterfall plots\\
4 & Excluded by spatial resolution test\\  
3 & Narrowband RFI identified in waterfall plots
\enddata
\end{deluxetable}

\section{Results}
The 10.5 hours of observations used by \citet{2015AJ....150..199T} were processed with a similar sensitivity to that obtained in the pilot study, but using data from both XX and YY polarisations. We confirm that no FRB-like signals were observed in that data set.  Our process identified two observations containing satellite tracks, and one heavily contaminated by RFI. The remaining 331 observations generated a single 7-sigma detection which when imaged was found to be in a sidelobe adjacent to a bright source. There was no detection corresponding to the most significant 6$\sigma$ detection highlighted in 2015. 

\begin{figure}[t]
\plotone{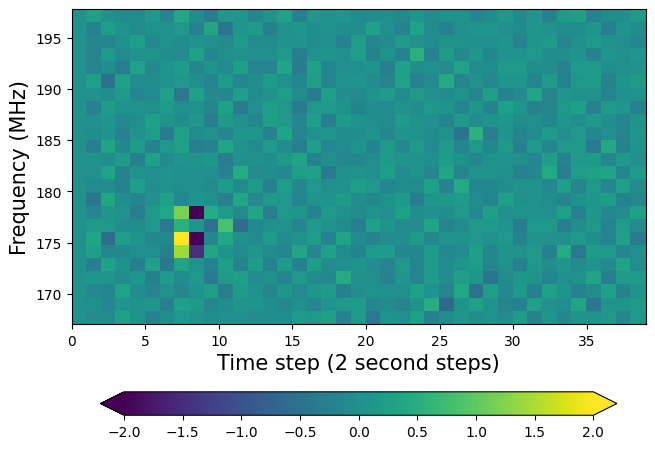}
\caption{Example waterfall plot showing narrowband RFI (Observation 1160580872). (177.5 MHz is a common DTV broadcast frequency in WA). Colour bar scale is in Jy/beam.
\label{fig:narrowband}}
\end{figure}

With the larger data set, the process described in the preceding section ended with three images which contained narrowband signals in a single pixel in the waterfall plots - see Figure \ref{fig:narrowband} as an example. These narrowband signals were also detected in the corresponding non-de-dispersed images.  Re-examination of images rejected at earlier stages provided a further 48 instances, some of which appeared more than once in the same 2-minute observation. They lay mostly between 171.5 to 180.5 MHz, although some extended outside that range.  

We tested these signals assuming they originate from digital TV (DTV) transmissions. In Western Australia there are 17 Digital TV stations broadcasting within our spectral window, which use specific frequencies in the range 177.5 to 198.6 MHz (\citet{ozdigitaltv_2024} using data from the ACMA\footnote{Australian Communications and Media Authority \url{https://www.acma.gov.au//}}). To test for aliasing of terrestrial sources on the horizon, whole-sky images were constructed for the relevant observations. These images indicated sky locations for the sources, suggesting that if they are DTV they are reflections of over-the-horizon sources. The random distribution of the signals in time (throughout the 3111 observations) and in the sky leads us to suggest these may be reflections from the ion trails of meteors. Such reflections are known to be detectable from DTV transmitters up to 2000 km from the telescope \citep{IMO_2024}, indicating that they could originate from any transmitter in Western Australia.

These narrowband signals had not been observed in the pilot (10.4 hr) dataset, but have proved numerous in the 87.6-hour search.

\section{Discussion}
\subsection{Assumptions and limitations}
In our work we were able to search 87.6 hours of data, from 107.1 hours of observations. This represents a loss of 18\% of the observation data set due to RFI and satellite contamination. In each snapshot image, approximately 1.5\% of the pixels are excluded by the scintillation mask, although being near-coincident with persistent radio sources these data are likely not usable for FRB detection.

Our method explicitly assumes effectively non-polarised sources; most linearly polarised signals will be excluded by the averaging of polarised images. With reference to Figure 3 of \citet{2022Sci...375.1266F}, due to lack of data it is not possible to rule out linear polarisation of FRBs in our spectral range. Therefore there is a possibility that detections near our 7$\sigma$ threshold may have been discarded.

\subsection{Upper bound on event rate}
We estimate the sensitivity of our search using a beam model for the MWA. A total of five pointing directions were used for the 3127 observations searched. An average of the beam models was calculated for each of the three data categories, weighted by the total observing time of each pointing direction, and corrected for pixels excluded by the scintillation mask. The cumulative fraction of the image area was computed for each decile of the normalised response, and these totals were multiplied by 7x the typical RMS noise to obtain a function of observing time.  For each sensitivity decile, division by the image area as a fraction of the full sky, gives event rate as a function of sensitivity.

Following the method of \citet{2020PASJ...72....3R}, Poisson statistics were used to obtain an upper bound on the event rate. A Poisson distribution is assumed, based on the assumption that for any given pixel, FRBs would be detected at random intervals in the time domain, with a constant average event rate $\lambda$ at a given sensitivity.

The probability of N detections in an observation of time t (measured in sky-days), is given by Equation \ref{eq:poisson1}. Setting N = 0 allows us to calculate the event rate for which a non-detection is due to chance (with probability p), given in Equation \ref{eq:poisson2}.

Event rates from the three data categories A, B, and C can be combined, for each fluence, via the equivalent total observation time (measured in sky-days), as given by Equation \ref{eq:poisson3}, which can be written as the reciprocal of the total event rate being the sum of the reciprocals of the category A, B, and C event rates (Equation \ref{eq:poisson4}.)

\begin{equation}
\label{eq:poisson1}
p=\frac{(\lambda t)^N e^{-\lambda t}}{N!}
\end{equation}

\begin{equation}
\label{eq:poisson2}
\lambda_{(N=0)} =\frac{-ln(p)}{t}
\end{equation}

\begin{equation}
\label{eq:poisson3}
\lambda =\frac{-ln(p)}{t_A + t_B + t_C}
\end{equation}

\begin{equation}
\label{eq:poisson4}
\frac{1}{\lambda} = \frac{1}{\lambda_A} + \frac{1}{\lambda_B} + \frac{1}{\lambda_C}
\end{equation}

\begin{figure}[h]
\plotone{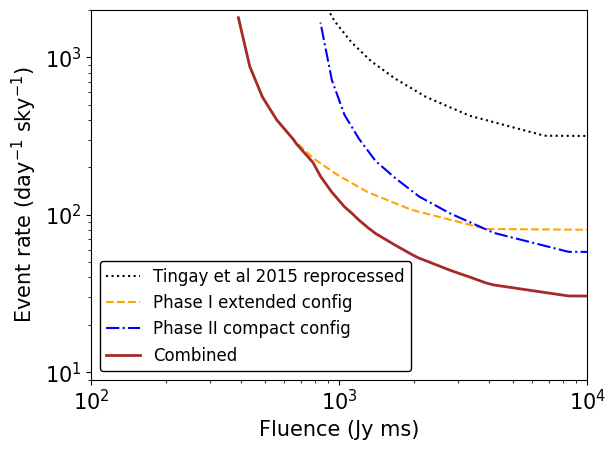}
\caption{Calculated upper bound event rates (DM=170 pc cm\textsuperscript{-3})
\label{fig:sky_rates}}
\end{figure}

\begin{figure}[h]
\plotone{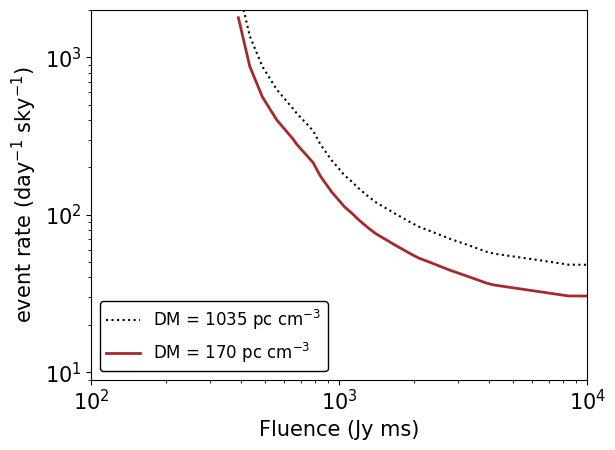}
\caption{Calculated upper bound event rates at lowest and highest DM values
\label{fig:sky_rates_DM}}
\end{figure}

The calculated event rates plotted in Figure \ref{fig:sky_rates} show the upper bounds, obtained from Equation \ref{eq:poisson2} with p = 0.05 (i.e. with 95\% confidence), given that no detection was made, together with the combined rate from Equation \ref{eq:poisson4}. Due to our method of processing data in units of 2-minutes of visibility data, the upper bound event rate varies with DM, and we plot upper bounds for the lowest and highest DM values tested, in Figure \ref{fig:sky_rates_DM}. Selected values are listed in Table \ref{table:selected_sky_rate}.

\begin{deluxetable}{lcc}[h]
\tablecolumns{3}
\tablecaption{Selected values of calculated upper bound event rates}
\label{table:selected_sky_rate}
\tablehead{
& \colhead{Event Rate at} & \colhead{Event Rate at} \\
\colhead{Fluence} & \colhead{DM=170 pc cm\textsuperscript{-3}} & \colhead{DM=1035 pc cm\textsuperscript{-3}} \\
\colhead{(Jy ms)} & \colhead{(sky\textsuperscript{-1} day\textsuperscript{-1})} & \colhead{(sky\textsuperscript{-1} day\textsuperscript{-1})}
}
\startdata
392 & 1783 & 2780\\
747 & 234 & 372\\
933 & 139 & 221\\
1680 & 64 & 102\\
3920 & 37 & 58\\
8400 & 31 & 48\\
\enddata
\end{deluxetable}

\begin{figure*}[t]
\plotone{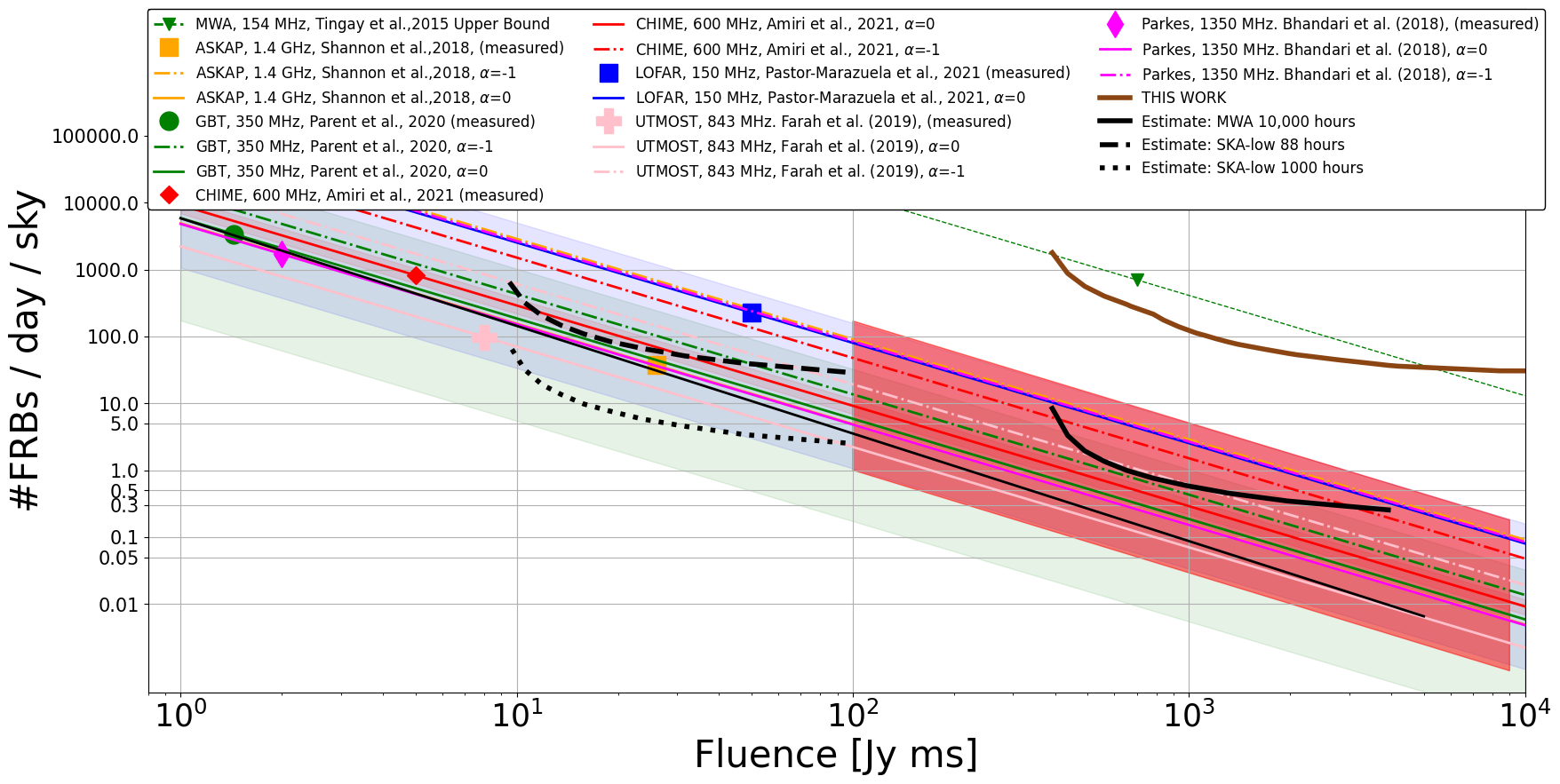}
\caption{Synthesis plot of \citet{2024arXiv240104346S} (reproduced with permission of the author) with, superimposed, upper bound from pilot study, upper bound from this study, and estimates of possible future searches using MWA and SKA.
\label{fig:synthesis_sky_rate}}
\end{figure*}

These upper bounds provide tighter constraints than the two blind search estimates given in Table \ref{table:MWAsearches}, but do not approach the expected rates calculated by \citet{2024arXiv240104346S}. Our upper bound is superimposed onto Figure 3 of \citet{2024arXiv240104346S} in our Figure \ref{fig:synthesis_sky_rate}, along with the upper bound determined in the pilot study.  The results of our search support these authors' extrapolation, and show that with our method a search with at least two orders of magnitude more data would be required to test their expectations.

\subsection{Future searches}
We calculated the sensitivity of our search method with an increased total observation time, assuming that the search used archival MWA data, restricted to extended configuration observations, and calibrated with hyperdrive - i.e. equivalent to our most sensitive 'category B'. A total of 10,000 hours of observation data (100x our search in this work) would be a serious test of \citet{2024arXiv240104346S} (see the chain-dot line of Figure \ref{fig:synthesis_sky_rate}). We believe our pipeline can limit the number of false positives to a manageable number, particularly by partial automation of the classification of false positives. Our classified detections (Table \ref{table:detections}) may be numerous enough to serve as a training set for a machine-learning based classifier.

It may be possible to improve the sensitivity further by mitigation of the texture which we attribute to confusion noise due to galactic plane setting (Sec. \ref{sec:imaging}). We recommend the development of a strategy to reject the narrowband detections which we attributed to DTV reflections from meteor trails. Alternatively, our method could be adopted as a method to catalogue these signals.

Our search was directed to 7$\sigma$ signals in pseudo-Stokes I snapshots. Due to the unquantified polarisation behaviour in the MWA frequency range, future researchers should consider searching for polarised signals, ie 7$\sigma$ signals in the XX and YY snapshot images, provided the overall sensitivity can be increased as indicated in the previous section.

We have calculated the sensitivity of a search using our method on data from SKA-Low \citep{2020SPIE11445E..12M}. We assume commensal observation as in the current study, and have used the expected RMS noise calculated by \citet{2022PASA...39...15S}. We obtain the dashed and dotted curves plotted on Figure \ref{fig:synthesis_sky_rate}, for 87 hours and 1000 hours of observation time. This indicates that a 87-hour search using SKA-Low would be slightly better than the proposed 10,000 hour search using the MWA, in terms of probing the expectations of \citet{2024arXiv240104346S}.

We have estimated the total computation requirements of three future search options, using today's costing at the commercial supercomputer used in our present work. The costs relate to imaging and FRB search - calibration is excluded on the assumption that we would be carrying out a commensal search. The estimates listed in Table \ref{table:future_searches} are based on imaging parameters relevant to \citet{2022PASA...39...15S}'s noise estimate.  It is to be expected that the compute cost may have decreased significantly by the time SKA data is available, but the node memory requirement may still be a challenge, suggesting a different search strategy or image size may be necessary. The putative MWA search, however, is possible with today's technology.

\section{conclusions}
We have demonstrated a viable method for searching archival data from the MWA telescope, which is computationally efficient and incorporates methods to deal with the known sources of false positives. 

We report no FRB detections, a result which is consistent with the expectations of \citet{2024arXiv240104346S}, obtained by an extrapolation of FRB observations at higher frequencies. However, our method is scalable to a data set two orders of magnitude greater, at which level the results may be a meaningful test of this model. 

Our processing approach has proven to be very sensitive to the detection of satellites and ionised meteor trails, and may prove to be a viable method to identify these either as sources of contamination or as objects of interest in the greater MWA archive.

We estimate that a search with today's technology using 10,000 hours of MWA data would be feasible and would be as sensitive as a search using 88 hours of SKA-Low data, at a similar cost in today's dollars.

\begin{deluxetable*}{rrrrc}[th]
\tablecolumns{5}
\tablecaption{Resource requirements for current work and estimates for some possible future searches}
\label{table:future_searches}
\tablehead{
\colhead{Search} & \colhead{\# observations} & \colhead{Node-hours} & \colhead{Cost (2024} & \colhead{Node RAM} \\
& & & \colhead{\$A prices)} & 
}
\startdata
Current Work & 3,127 & 1,564 & \$1,329 & 84 GB\\
10,000 hr MWA & 355,000 & 178,000 & \$151,000 & 84 GB\\
88 hr SKA-low & 3,127 & 225,000 & \$191,000 & 12 TB\\
1000 hr SKA-low & 35,500 & 2,560,000 & \$2,170,000 & 12 TB\\
\enddata
\end{deluxetable*}

\begin{acknowledgments}
\section{Acknowledgements}
The authors thank CSIRO, DUG Technology and Curtin University for supporting this work financially and in kind.

We also thank the ICRAR EoR team for provision of the hyperdrive-calibrated observations used in our work.

The scientific work made use of Inyarrimanha Ilgari Bundara, the CSIRO Murchison Radio-astronomy Observatory. We acknowledge the Wajarri Yamaji people as the traditional owners of the Observatory site. This work was supported by resources provided by the Pawsey Supercomputing Centre with funding from the Australian Government and the Government of Western Australia.
\end{acknowledgments}

\section{ORCID IDs}
Ian Kemp \url{https://orcid.org/0000-0002-6637-9987} \\
Steven Tingay \url{https://orcid.org/0000-0002-8195-7562} \\
Daniel Mitchell \url{https://orcid.org/0000-0002-1828-1969}

\bibliography{KempEtAl}{}
\bibliographystyle{aasjournal}
\end{document}